\documentclass[twocolumn,prl,superscriptaddress,aps,showpacs]{revtex4}

\usepackage{graphicx}
\usepackage{color}
\usepackage{dcolumn}
\usepackage{amsmath}

\newcommand{\vc}[1]{\mbox{\boldmath $#1$}} 
\newlength{\figwidth}
\newlength{\figwidthb}
\newcommand{\eb}{EuB$_{6}$} 

\begin{document}

\title{Spin-Split Conduction band in \eb\ and Tuning of Half-Metallicity with External Stimuli}
\author{Jungho Kim}
\address{Advanced Photon Source, Argonne National Laboratory,
Argonne, Illinois 60439, USA}
\address{Department of Physics,
University of Toronto, Toronto, Ontario M5S~1A7, Canada}
\author{Wei Ku}
\address{Condensed Matter Physics and Materials Science Department, Brookhaven National Laboratory, Upton, New York 11973, USA}
\author{Chi-Cheng Lee}
\address{Condensed Matter Physics and Materials Science Department, Brookhaven National Laboratory, Upton, New York 11973, USA}
\address{Institute of Physics, Academia Sinica, Nankang, Taipei 11529, Taiwan}
\author{D. S. Ellis}
\address{Department of Physics, University of Toronto, Toronto,
Ontario M5S~1A7, Canada}
\author{B. K. Cho}
\address{Center for Frontier Materials and Department of
Materials Science and Engineering, GIST, Gwangju 500-712, Korea}
\author{A. H. Said}
\address{Advanced Photon Source, Argonne National Laboratory,
Argonne, Illinois 60439, USA}
\author{Yu. Shvyd'ko}
\address{Advanced Photon Source, Argonne National Laboratory,
Argonne, Illinois 60439, USA}
\author{Young-June Kim}
\email{yjkim@physics.utoronto.ca} \address{Department of
Physics, University of Toronto, Toronto, Ontario M5S~1A7, Canada}

\date{\today}

\begin{abstract}

We report Eu L$_3$-edge resonant inelastic x-ray scattering (RIXS)
investigation of the electronic structure of EuB$_6$. We observe
that the RIXS spectral weight around 1.1 eV increases dramatically
when the system is cooled below the ferromagnetic ordering
temperature and follows the magnetic order parameter. This spectral
feature is attributed to the inter-site excitation from the local
4$f$ orbital to the spin-split 5$d$ orbital on the neighboring site,
illustrating the essential role of exchange splitting of the
conducting electrons. Based on our density functional theory
calculations, the RIXS data suggest that EuB$_6$ at low temperature
can be consistently described with a semi-metallic electronic
structure with incomplete spin-polarization. We propose routes to
achieve half-metallicity in \eb, which utilize the strong tunability
of the electronic structure against gate voltage, strain, and
magnetic field.
\end{abstract}


\maketitle

Half-metals are characterized by fully spin-polarized charge
carriers, and can be very useful for spintronic applications. One of
the materials that have attracted much attention as a potential
half-metal is \eb~\cite{kunes}. In \eb, a ferromagnetic (FM) order
sets in below ~15 K, as shown by neutron
scattering~\cite{Henggeler98}, magnetization~\cite{sullow98}, and
specific heat~\cite{sullow00} measurements. Its electronic property
is strongly correlated with the FM order, as indicated by the sharp
drop in resistivity ($\rho$), the blue-shift of the unscreened
plasma frequency ($\omega_p$) and the colossal magneto-resistance
(CMR), all occurring in the vicinity of the FM ordering
temperature~\cite{Guy80,degiorgi}. In a recent local density
approximation (LDA)+U calculation, it was suggested that a
substantial conduction band (CB) and valence band (VB) exchange
splitting could lead to a half-metallic ground state in \eb\
\cite{kunes}. However, a recent Andreev reflection
spectroscopy~\cite{Zhang08} reported that only about half of the
carriers are spin polarized, which indicates that only the VB is
spin-split in the ground state with almost no CB splitting. Indeed,
an angle resolved photoemission (ARPES) experiment has reported a
splitting of the VB, but does not address the conduction band
issue~\cite{denlinger2}.

Resonant inelastic x-ray scattering (RIXS) is a bulk-sensitive and
element-specific probe of electronic structure that can address the
issue of CB spin-splitting in \eb. Much attention has been paid to
transition metal K-edges in the hard x-ray regime to uncover various
momentum-dependent
excitations~\cite{Hasan00,Yjkim02,Grenier05,Jphill08,Ellis10}. Full
instrumental capability of the transition metal $K$-edge RIXS can be
extended to the $L_3$-edges of rare-earth (RE) elements
straightforwardly, since they are in the same energy range as the
transition-metal K-edges. Until now, however, applications of hard
x-ray RIXS to study RE compounds have been limited to core
excitations~\cite{Yamaoka06}. At RE $L_3$-edges, RIXS probes $5d$
states directly, enabling one to study $5d$ related excitations. In
particular, the $4f-5d$ transitions in RE compounds provide
information about the interaction between local moments and
conduction electrons. Since partially occupied 4$f$ states are
responsible for local magnetic moments, and delocalized $5d$ states
form the conduction band, magnetic properties of these materials
crucially depend on the inter-site interaction of the 4$f$ moments
through the intra-atomic $d$-$f$ exchange mediated by the conduction
electrons. In addition, such an interaction between the 4$f$ moments
through the $d$-$f$ exchange mediated by conduction electrons often
leads to a diverse range of magnetic and electronic properties, such
as CMR and
half-metallicity~\cite{Wolf01,Coey04,deGroot83,Katsnelson08,Schmehl07}.

In this Letter, we report Eu L$_3$-edge RIXS study of the electronic
structure of EuB$_6$ at low temperatures. We observe a large
resonant enhancement of excitations in the energy range of 1-6 eV,
which shows a complex incident energy dependence. When the incident
photon energy is fixed at $E_{i}$=6982~eV, we could selectively
enhance low energy excitations below 2 eV, which correspond to
transitions from the local $4f$ orbitals to the spin-split 5$d$
orbitals on the neighboring sites. This low energy RIXS spectral
weight grows significantly as the sample is cooled below the FM
transition temperature, closely mimicking the temperature dependence
of the ordered magnetic moment. This observation suggests that the
FM ordering is accompanied by a fairly substantial exchange
splitting of the conduction band, which opens up the $4f-5d$
transition channel. Our density functional theory calculations
combined with the RIXS findings point to a slightly doped
semi-metallic system with incomplete spin-polarization. This
provides a natural way to explain the previously reported Andreev
reflection data~\cite{Zhang08}. We also show that such a system can
be driven into a desired half-metallic state via gate voltage,
strain, or magnetic field.

$Experiment$ - The Eu L$_3$-edge RIXS measurements were performed
using the MERIX spectrometer at the 30ID beamline of the Advanced
Photon Source (APS). The single-crystal sample was synthesized by a
boro-thermal method as described in detail elsewhere\cite{rhyee2}.
The sample is mounted in a closed-cycle He cryostat. Throughout the
RIXS measurements, sample temperature was controlled within 0.2~K of
stability. The total energy resolution of the MERIX spectrometer at
the Eu L$_3$-edge is $140$~meV, determined by the
full-width-half-maximum (FWHM) of the elastic scattering peak. This
is achieved using a 1~m Ge$(620)$ spherical, diced analyzer, and a
position-sensitive microstrip detector~\cite{HVA05}. The incident
photon polarization component was perpendicular to the scattering
plane ($\sigma$-polarization).

%
%

\begin{figure}[t]
\hspace*{0.07cm}
\centerline{\includegraphics[width=3.4in,angle=0]{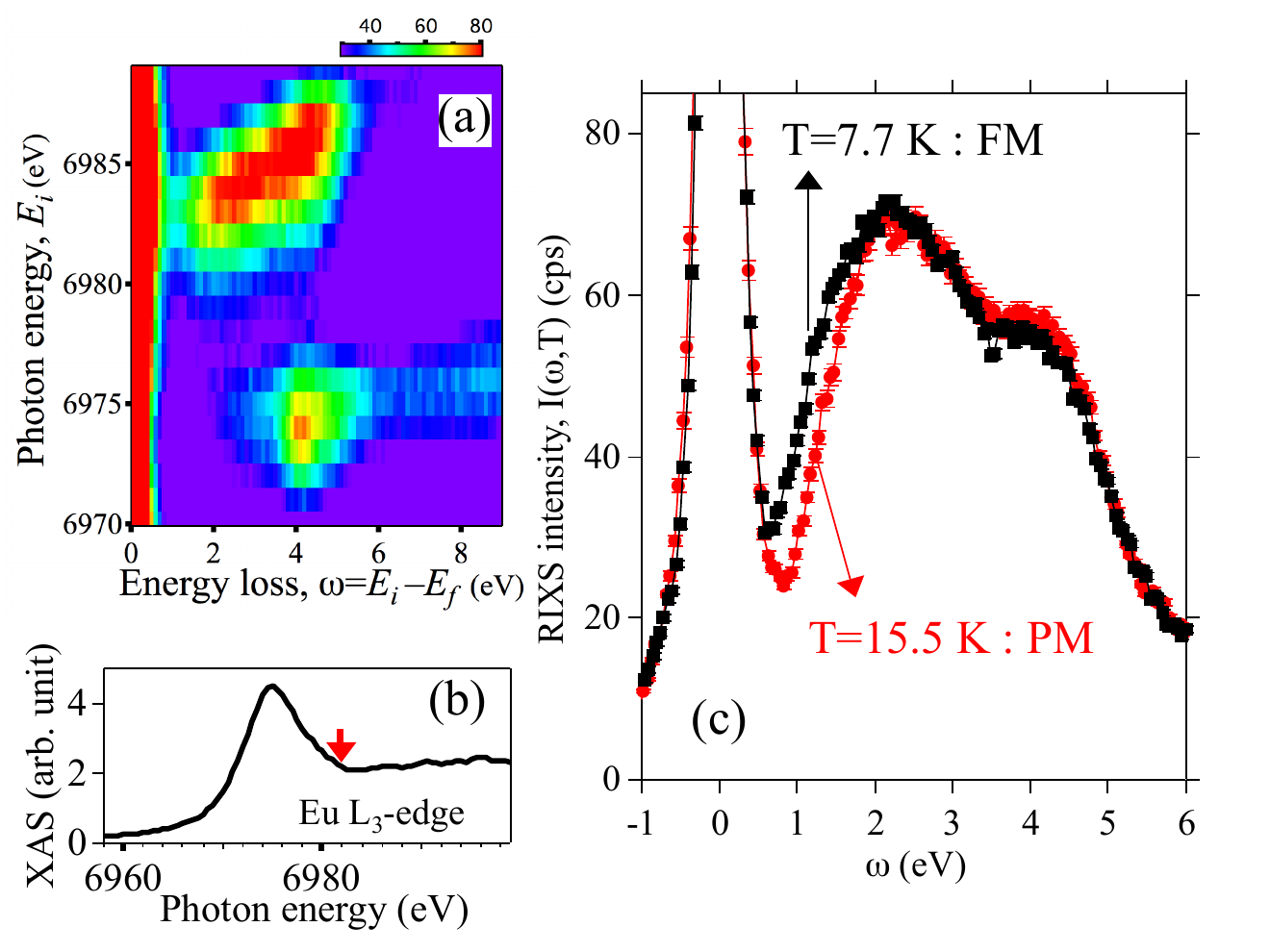}}
\vspace*{-0.9cm}
\caption{(Color online) (a) Color map of RIXS intensity as a
function of energy loss $\omega$ and incident photon energies
$6970~{\mathrm{eV}}\le E_i \le 6989$~eV. (b) The Eu L$_3$-edge XAS
spectrum measured in the partial fluorescence yield. The arrow
indicates $E_i$=6982~eV which is used for the $T$ dependence study.
(c) Comparison of the RIXS spectra in the FM phase ($T$=7.7 K) and
the PM phase ($T$=15.5 K) for $E_i$=6982~eV.}\label{fig:resonance}
\end{figure}

Figure~\ref{fig:resonance}(a) shows the incident photon energy,
$E_{i}$, dependence of the RIXS spectra at the momentum transfer of
$\vc{Q}$=(1.5,0,0)~\cite{selfcor}. Each spectrum is measured by
scanning the scattered photon energy, $E_f$, with $E_i$ fixed, and
plotted as a function of the energy loss, $\omega \equiv E_i-E_f$.
The observed resonance enhancement of the spectral features is quite
large, and comparable to that in strongly correlated copper
oxides~\cite{Yjkim02}. A well-defined spectral feature at $\omega
\approx 4$~eV is found to resonate at around $E_{i}$ = 6975~eV,
which corresponds to the peak in the Eu L$_3$ x-ray absorption
spectrum (XAS) shown in Figure~\ref{fig:resonance}(b). Although
there is only one prominent XAS peak at 6975~eV, there are
additional transitions hidden in the broad, featureless intensity at
higher energy. Since these XAS final states constitute the
intermediate states in our RIXS experiment, we can pick out these
additional intermediate states from the RIXS spectra. The $\omega
\approx 4$~eV feature seems to resonate again around $E_{i}$ =
6985~eV, while there is a significant low energy intensity around
$\omega =2$~eV, which resonates around $E_i=$6982~eV. A detailed
study of the $E_i$ dependence of RIXS spectrum will be presented
elsewhere, and we only focus on the temperature and momentum
dependence of this low energy excitation here.

%
%

\begin{figure}[t]
\vspace*{0.1cm}\centerline{\includegraphics[width=3.3in,angle=0]{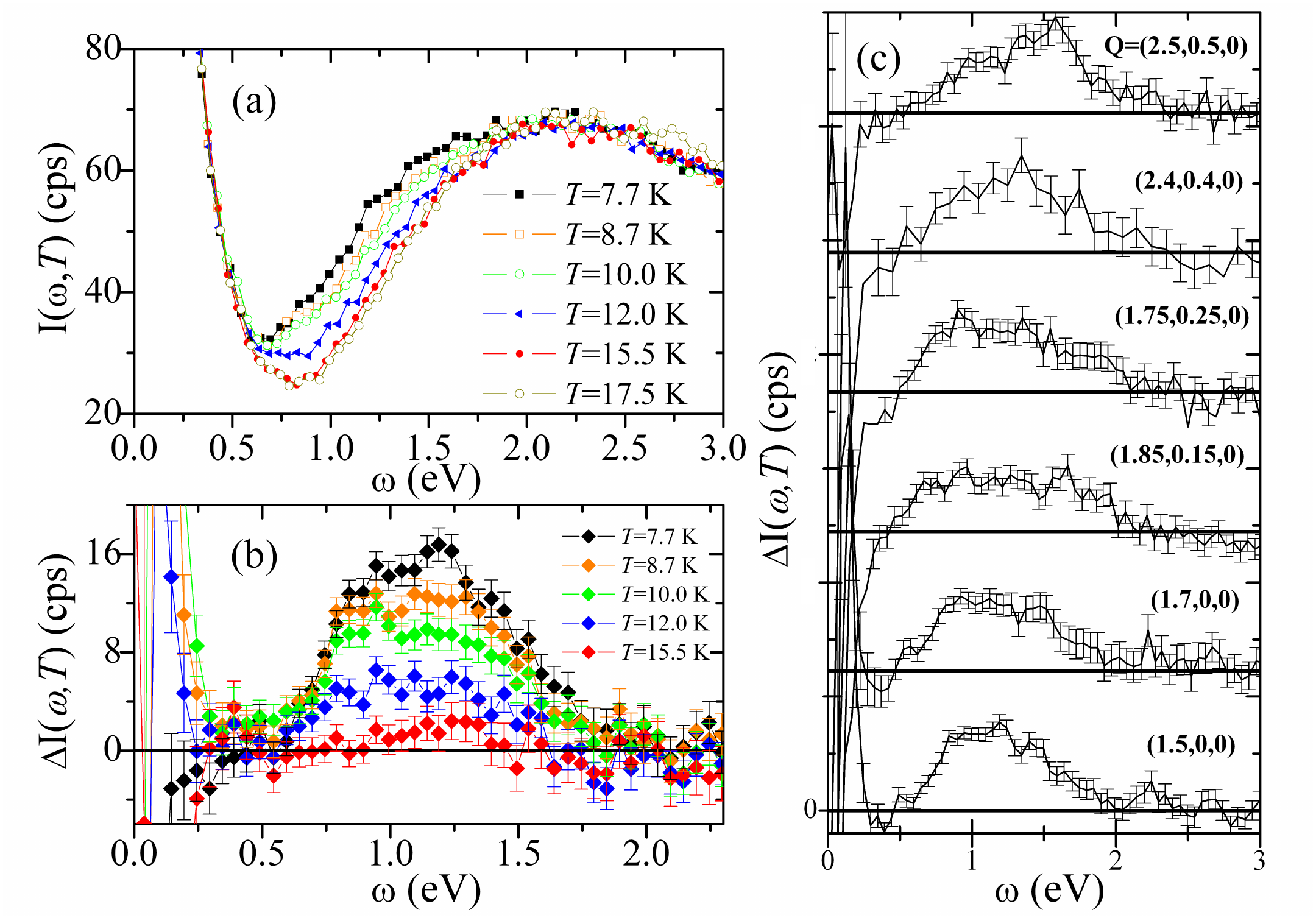}}
\vspace*{-0.4cm}%
%
\caption{(Color online) (a) RIXS spectra for $E_{i}$=6982~eV at
various temperatures from above to below $T_{c1}=15~K$. The RIXS
intensity below 2~eV grows systematically as $T$ decreases below
$T_{c1}$. (b) The difference between the FM phase and the PM phase:
$\Delta$I$(\omega,T)$=I$(\omega,T)-$I$(\omega,17.5\rm{K})$. (c)
$\Delta$I$(\omega,7.7~K)$ for different $\vc{Q}$. Top three panels
corresponds to the $M-\Gamma$ direction, while the bottom three are
obtained along the $\Gamma-X$ direction.}\label{fig:tempdep}
\end{figure}

In Fig.~\ref{fig:resonance}(c), the RIXS spectra obtained in the FM
phase and the PM phase are compared. The two RIXS spectra are almost
identical except for the region between 0.5~eV and 2~eV. Note that
the difference in the 0.5-2~eV region cannot be due to a shift,
since the spectra overlap both below and above this frequency
region. The quasi-elastic background, which usually varies with
temperature, stays the same, since actual temperature change is
quite small. Clearly the RIXS intensity in the FM phase in the
0.5-2~eV region is significantly greater than that in the PM phase.
Detailed temperature dependence of the RIXS spectra below and above
the FM transition are plotted in Fig.~\ref{fig:tempdep}(a). In order
to show the temperature dependence more clearly, we plot the
background subtracted spectra in Fig.~\ref{fig:tempdep}(b), in which
the spectrum obtained at $T=17.5$~K is used as a background. As
temperature is lowered, RIXS intensity grows gradually without
changing its energy position. We have also obtained difference
spectra between the FM and PM phase for a number of $\vc{Q}$
positions which correspond to momentum transfers along the
$\Gamma-X$ and $\Gamma-M$ of the Brillouin zone in the reduced zone
scheme. Figure~\ref{fig:tempdep}(c) shows that the FM induced RIXS
intensity exhibits little momentum dependence.

\begin{figure}[t]\hspace*{-0.15cm}
\vspace*{-0.2cm}\centerline{\includegraphics[width=3.4in,angle=0]{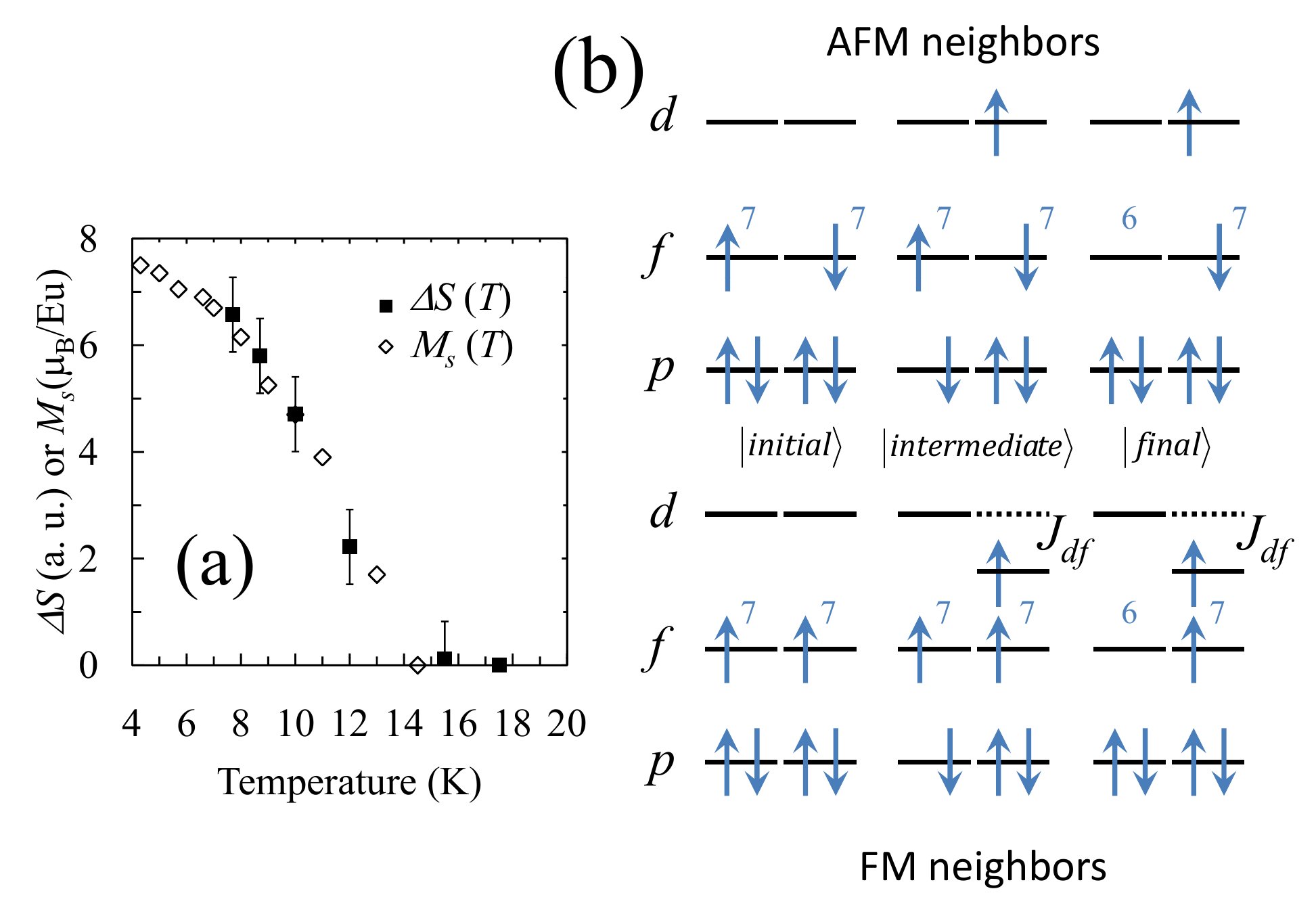}}
\vspace*{-0.7cm}%
%
\caption{(Color online) (a) Filled squares: The RIXS integrated
intensity change, $\Delta S(T) \equiv \int{\Delta I (\omega,T)
d\omega}$ ($0.5 \le \omega \le 2$~eV). Open diamonds: Spontaneous
magnetic moment per Eu atom obtained from the Arrott plot analysis
of the magnetization data (from Ref.~\cite{sullow00}). (b)
Schematics of the RIXS processes when magnetic moments on
neighboring atoms are opposite (top figure) or parallel (bottom).
The numbers next to the $f$-state spins indicate the number of
electrons in the corresponding $f$-state. $J_{df}$ is the effective
on-site $d$-$f$ exchange coupling.}\label{fig:quantity}
\end{figure}

In Fig.~\ref{fig:quantity}(a), the integrated RIXS intensity from
Fig.~\ref{fig:tempdep}(b) is compared to the temperature dependence
of the spontaneous magnetic moment from Ref.~\cite{sullow00}, which
represents magnetic order parameter. The close similarity between
the two indicates that the RIXS spectral weight increase upon
entering the FM phase is intimately related to the FM order
parameter. Such a strong temperature dependence of the RIXS spectra
has not been observed before, since RIXS typically measures in the
energy range much higher than thermal energy. Our observation is
reminiscent of the case in manganites \cite{Grenier05}, for which
the spectral weight of the inter-site $d$-$d$ excitation was found
to be sensitive to the magnetic ground state, but quantitative
comparison of the magnetization and the RIXS intensity was not
possible.

%
%

\begin{figure}[t]\hspace*{-0.2cm}
\vspace*{-0.2cm}\centerline{\includegraphics[width=3.5in,angle=0]{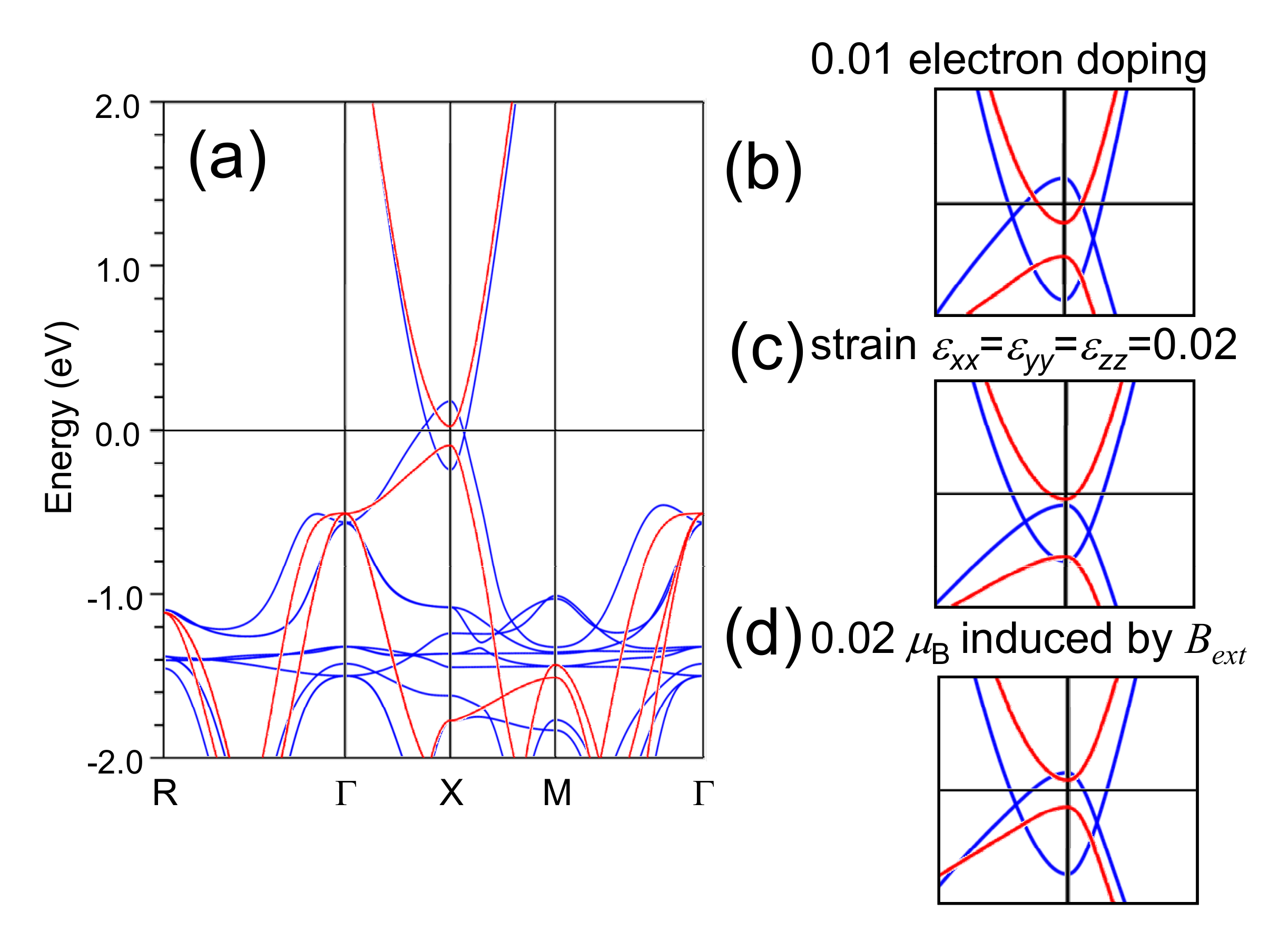}}
\vspace*{-0.4cm}%
%
\caption{(Color online) (a) LDA+U band structure of \eb\ with 0.5~eV
self-interaction correction. The metallic spin-majority and gapped
spin-minority bands are colored blue and red, respectively. (b) Band
structures near the X point under 0.01 electron doping per formula
unit. This represents the realistic band structure of our sample at
low temperatures. Applied gate voltage could drive the system into
the half-metallic state shown in part (a). Likewise, the effects of
(c) stress and (d) external magnetic field can be used to tune the
electronic structure of \eb.}\label{fig:band}
\end{figure}

$Theory$ - To facilitate analysis of our excitation spectra, we
performed a first-principles density functional theory calculation
of the electronic structure within the LDA+U approximation.  To
remove the well-known self-interaction problem of the approximate
energy functional, we apply a ``real-space scissor operator'' to
split the energy of the occupied and unoccupied orbitals by another
0.5 eV in the low-energy effective Hamiltonian obtained from the
first-principles Wannier function analysis~\cite{Weiku02}.  This
procedure, unlike the rigid band shift produced by the common
``$k$-space scissor operator'', allows $k$-dependent
re-hybridization of the orbitals, thus producing a different band
dispersion and a set of consistent eigen-orbitals. As shown in
Fig.~\ref{fig:band}(a), in agreement with previous
calculations~\cite{massida,kunes}, the band structure of \eb\ around
the Fermi level ($E_F$) is characterized by the occupied B-2$p$ VB,
Eu-4$f$ lower Hubbard band, and the unoccupied Eu-5$d$ CB hybridized
with B-2$p$. The VB and CB approach each other in the vicinity of
the X point of the Brillouin zone, while the flat 4$f$ band stays
about 1-2~eV below $E_F$~\cite{Takakuwa78}.

We associate the observed RIXS excitation in the 0.5-2 eV range with
the transition from the $4f$ bands to the bottom of the conduction
($5d$) bands near the X point. It is well-known that the low energy
particle-hole excitations below 5~eV are dominated, especially at
the Eu ion, by the transitions from the $4f$ to the $5d$
states~\cite{Kimura92,Singh07,Jun08}. Magneto-optical Kerr effect in
fact identified electronic transitions involving the localized 4$f$
electron states in the same energy range~\cite{caimi}. In addition,
our observation of very little momentum dependence suggests that the
excitation involves relatively flat bands (e.g., $f$-bands), since
excitations from wide bands tend to show strong momentum
dependences. We find that the inclusion of spin-orbit (SO) coupling
(results not shown) broadens the 4$f$-bandwidth by 0.5~eV, which can
account for the experimentally observed width of this spectral
feature.

%
%

The observed high sensitivity of the RIXS spectral weight to the FM
order could result from $inter$-$site$ $d$-$f$ excitations. The
leading RIXS process for such an inter-site $d$-$f$ excitation is
illustrated schematically in Fig.~\ref{fig:quantity}(b). In the
intermediate state, the core electron is kicked directly to the
$5d$-orbital of the neighboring site. Due to the inter-site nature
of the process, it has a small matrix element and requires higher
incident energy, but it is more sensitive to the on-site magnetic
correlation~\cite{others}. The FM order-induced changes in the
inter-site $d$-$f$ excitations arise from the effective on-site
$d$-$f$ exchange, $J_{df}$, which lowers the $f-d$ transition energy
between FM neighbors as depicted in Fig.~\ref{fig:quantity}(b). On
the other hand, when neighbors are anti-ferromagnetically aligned,
only the higher energy $f-d$ transitions could occur.  The size of
$J_{df}$ can be estimated from our LDA+U calculations. Since the
energy difference between the spin-up and spin-down Eu 5$d$ states
(CB) is about 0.6~eV and there are seven $f$ electrons, $J_{df}$ is
about 0.08~eV. That is, the finite $J_{df}$, i.e., the spin-split CB
in the FM phase could give rise to the observed temperature
dependence of the RIXS spectral weight. This illustrates the
importance of the spin-polarization of the CB.

Our experimental and theoretical results, as well as the previous
observation of ~56$\%$ carrier spin polarization~\cite{Zhang08}, can
be consistently described by assuming that the system is slightly
($\sim$1$\%$) electron doped, presumably from some (3$\%$) B
vacancies~\cite{Monnier01}.  As shown in Fig.~\ref{fig:band}(b), due
to the efficient $d$-$f$ exchange in the FM phase, the
low-temperature electronic structure would be that of a semi-metal
with incomplete spin-polarization, consisting of electron pockets of
both spin, and a hole pocket only of the majority spin.  Such an
electronic structure is expected to be highly tunable against
external stimuli.  For example, Figs.~\ref{fig:band}(a), (c), and
(d) demonstrate how half-metallicity can be achieved via gate
voltage, tensile strain, and external magnetic field, respectively.
Changes in the band dispersion, as well as the exchange splitting,
give rise to such tunability towards half-metallicity; effectively,
the decrease of the minority-spin conduction band dispersion leads
to a half-metallic band structure. Such a sensitive response to
external stimuli makes \eb\ an interesting potential candidate for
spin-dependent transport devices, exploiting the spin-filter, giant
magnetoresistance, or tunneling magnetoresistance effects.

To summarize, our Eu L$_3$-edge resonant inelastic x-ray scattering
(RIXS) experiments have revealed magnetically sensitive RIXS
spectral weight, which can be understood as arising from the
lowering of the inter-site $f-d$ transition energy between
ferromagnetic neighbors by the effective on-site $d$-$f$ exchange.
Our density functional theory calculations combined with the RIXS
findings suggest an incomplete spin-polarized semi-metallic
electronic structure for \eb\ at low temperature. We also find that
highly desirable half-metallicity can be achieved in such an
electronic structure, which  exhibits a strong tunability against
gate voltage, strain, and magnetic field. It will be interesting to
study thin-films~\cite{bachmann70,Dorneles04} and
nanowire/nanotube~\cite{Xu07} of \eb, which provide a platform for
tuning the electronic and magnetic properties via these external
stimuli.

\acknowledgements{Work at U of Toronto was supported by the NSERC of
Canada, Canadian Foundation for Innovation, and Ontario Ministry of
Research and Innovation. Theoretical work at Brookhaven was
supported by the U. S. DOE, Office of Science Contract No.
DE-AC02-98CH10886. Use of the Advanced Photon Source was supported
by the U. S. DOE, Office of Science, Office of Basic Energy
Sciences, under Contract No. W-31-109-ENG-38.}

\bibliography{eubrixs}

\begin{thebibliography}{34}
\bibitem[{oth()}]{others}
\bibinfo{note}{We note that there are many other RIXS processes that are
  responsible for other strong spectral features at higher energies, but only
  magnetically sensitive processes are considered here.}
\end{thebibliography}

\end{document}